# Boron Nitride Nanotubes as Efficient Surface Absorbers for Air Pollutant Gas Molecules: Insights from Density Functional Theory

Chaithanya Purushottam Bhat[a], Joy Mukherjee[b], Antara Banerjee[c] and Debashis Bandyopadhyay[*a]

[a]Department of Physics, Birla Institute of Technology & Science, Pilani, Rajasthan – 333031

[b]Human-Computer Interaction Institute, Carnegie Mellon University

Pittsburgh, Pennsylvania, United States

[c]Science Department, Vidya Niketan Birla Public School Pilani, Rajasthan, 333031, India

[*]e-mail: Debashis.bandy@gmail.com

**Abstract**

This study investigates into the adsorption sensing capabilities of single-walled (5,5) boron nitride nanotubes (BNNTs) towards environmental pollutant gas molecules, including $CH_2$, $SO_2$, $NH_3$, $H_2Se$, $CO_2$ and $CS_2$. Employing a linear combination of atomic orbital density functional theory (DFT) and spin-polarized generalized gradient approximation (GGA), the investigation reveals the nanotube's robust adsorption behavior without compromising its structural integrity. Thermodynamic and chemical parameters, such as adsorption energy, HOMO-LUMO gap, vertical ionization energy, and vertical electron affinity, highlight the (5,5) BNNTs' potential as efficient absorbents for pollutant molecules. Infrared spectroscopy confirms the formation of distinct BNNT-gas complexes. These findings underscore the promising application of BN nanotubes as absorbents for common gaseous pollutants, essential for developing sensors to enhance indoor air quality.

## 1 INTRODUCTION

Boron Nitride Nanotubes (BNNTs) have garnered attention for their potential as effective gas molecule absorbents, owing to their distinctive physical and chemical properties. With a notable high surface area and the ability to be tailored for specific chemical characteristics, BNNTs stand out as appealing candidates for gas adsorption applications. Furthermore, their exceptional thermal and chemical stability makes them well-suited for deployment in challenging environments. Numerous studies have explored the versatility of BNNTs as gas molecule absorbents, encompassing gases like hydrogen, methane, and carbon dioxide. For instance, one study demonstrated the selective adsorption of $CO_2$ by BNNTs functionalized with amine groups. In contrast, another study highlighted BNNTs' efficiency in adsorbing hydrogen at room temperature. Since the discovery of single-walled carbon nanotubes (SWCNTs) by Iijima [1], these novel materials have found diverse applications in areas such as hydrogen storage, chemical sensors, and electronic devices [2–11]. Recently, nanotubes have garnered significant interest in gas sensor applications [7–9]. Various nanoscale chemical gas sensors based on carbon nanotubes have been employed to detect minimal quantities of different gas molecules [10, 11]. However, the sensitivity of such sensors depends largely on the SWCNT tubular diameter and chirality, which can exhibit metallic or semiconducting behavior. Hence, the primary focus is on identifying alternatives with different physical and chemical properties. Computational modeling, aimed at finding materials that can serve as ideal substitutes for SWCNTs, has spotlighted boron nitride nanotubes (BNNTs). Despite sharing similar morphology with CNTs, BNNTs exhibit distinct chemical properties. Their high surface area, tunable chemical characteristics, and resilience to harsh environments make them attractive for gas adsorption. Numerous studies have investigated their potential for absorbing various gases, such as hydrogen, methane, and carbon dioxide.

For biomedical applications, BNNTs have been introduced as potential tools for adsorption due to their high surface bonding energy, as reported [15–19]. Unlike carbon nanotubes, the semiconducting nature of BNNTs remains consistent regardless of their diameter and chirality [20,21]. With high chemical inertness and structural stability, BNNTs pose no hazards to health and the environment, rendering them versatile for a broad range of applications. Current reports demonstrate the efficient chemical adsorption of different pollutant gas molecules by pure BNNTs [22,24]. Therefore, it is crucial to comprehend the chemical and physical properties of BNNTs for improving their suitability as adsorbents for common gas pollutants like $CH_2$, $SO_2$, $NH_3$, $H_2Se$, $CO_2$ and $CS_2$ on the external surface of (5, 5) single-walled boron nitride nanotubes (BNNTs). Initial tests indicate the superior performance of bare BNNTs over defect-induced BNNTs. Density functional theory (DFT) calculations at the B3LYP level are conducted to gain insights into the structural properties, chemical reactivity, adsorption energy, and more of BNNTs.

## 2 COMPUTATION METHODOLOGY

In this report, the first principle calculations of all the geometry optimizations and geometric structures are performed using Gaussian'09 software at the level of density functional theory (DFT) with B3LYP functional and 6-311++G(d,p) basis set [25–27] inbuilt in the software. We begin with studying the pristine (5,5) single-walled BNNT molecule and analyzing its electrostatic and molecular orbital (MO) properties before performing the energetic study. We terminate the ends of the optimized BNNT structure using Hydrogen atoms to model a repeating, more comprehensive system to work with. We have simulated and visualized the Molecular Orbital configurations and Electrostatic potential on the Iso-electron density surface using DFT analysis with B3LYP/ 6-311++G(d) basis set for all calculations because of its suitability with all atoms involved in the work. We begin our investigations by constructing optimized structures of BN nanotubes complexed with pollutant molecules of interest. Atoms were strategically free in space to avoid any restriction during simulations. In the computational model, the pollutant gas molecule adsorption occurs on (5,5) BNNT containing 45 boron and 45 nitrogens with two sides terminated by 20 hydrogen atoms are investigated for different orientations of the gas molecules on the BNNT surface. The system was fully relaxed (keeping all position coordinates free) until the following convergence criteria were achieved: $10^{-6}$ Harte for the total energy.

## 3 RESULTS AND DISCUSSIONS

### 3.1. Global reactivity descriptors

Optimized pure and gas molecule adsorbed (5,5) pristine BN nanotube is shown in Fig. 1. To understand the physical and chemical properties of pure and gas molecule adsorbed BNNTs, we have calculated several energetic parameters associated with the optimized structures following our previous reports on different systems [26-49]. Following the reports [26-49], HOMO and LUMO can be taken as the measure of (electron donor), and LUMO (electron acceptor) energies are essential parameters. The energy gap ($E_{LUMO}$-$E_{HOMO}$) is the essential chemical parameter that illustrates molecules' chemical reactivity and thermodynamic stability. In other words, more charge polarization and reactive molecules are responsible for the small HOMO-LUMO energy gap. Therefore, in the present work, lower values of the HOMO-LUMO gap may help to adsorb the gas molecule over the BNNT surface through the exchange of electrons. The DFT method defined chemical parameters, such as HOMO and LUMO gap (Δ), vertical ionization potential (IP) and electron affinity (EA), chemical potential (μ), chemical hardness (η), chemical softness (S), electronegativity (χ) and electrophilicity index (ω) by equations given below. As a basic calculation of these chemical parameters, first, we have calculated electron affinity (EA) and the ionization potential (IP). For the 'N' electron system with total energy E, chemical potential (μ) and chemical hardness (η) are defined by the 1st and 2nd-order derivatives, respectively, as follows:

$$\mu = \left( \frac{\delta E}{\delta N} \right)_{V(\vec{r})} = -\chi \text{ and } \eta = \frac{1}{2} \left( \frac{\delta \mu}{\delta N} \right)_{V(\vec{r})} = \left( \frac{\delta^2 E}{\delta N^2} \right)_{V(\vec{r})} ----(1)$$

Following Koopmans' approximation [50] using energies of HOMO (EH) and LUMO (EL) orbitals electronegativity, different chemical parameters, such as HOMO-LUMO gap ($E_g$), ionization potential (IP), electron affinity (EA), chemical potential (μ), chemical hardness (η), chemical softness (S) and electrophilicity index (ω) are defined as follows:

$$E_g = E_H - E_L;\ IP = E_{HOMO};\ EA = E_{LUMO}; \mu = \frac{1}{2}(IP + EA) = -\chi; \eta = \frac{1}{2}(IP - EA); S = \frac{1}{2\eta}; \omega = \frac{\mu^2}{2\eta}$$

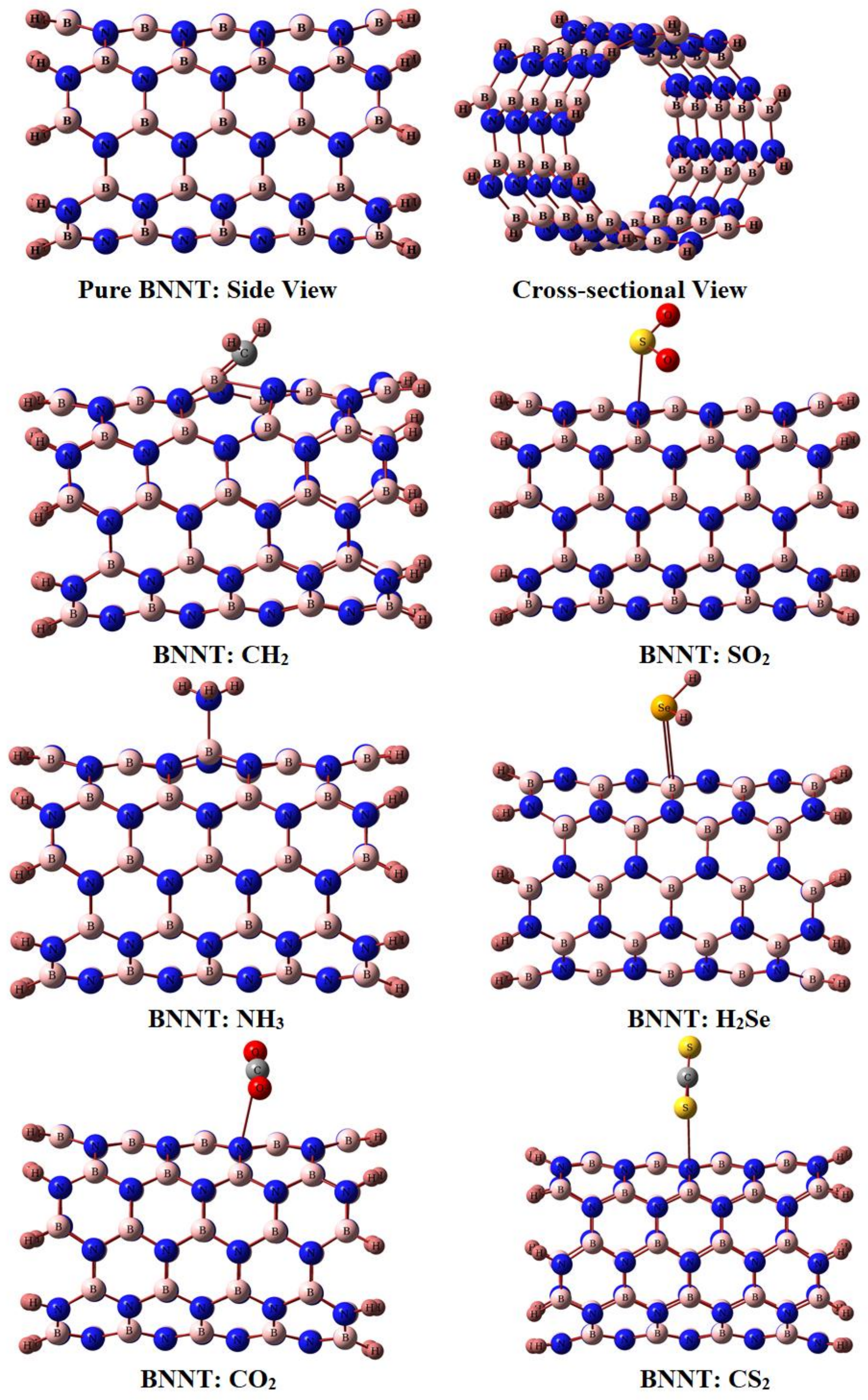

*Figure 1: Optimized pure and pollutant gas molecule adsorbed BNNT structure*

The chemical Potential μ measures the tendency to leave the system of an electron. Since it is associated with molecular electronegativity, a higher value of μ indicates that it is more challenging to lose an electron but easier to gain one. The electron affinity (EA) parameter represents the efficiency of the molecules/compound in attracting electrons. Chemical hardness (η) and softness (S) are essential chemical parameters to understand the behavior of chemical properties of the systems. When a molecule or a compound is chemically hard, the meaning is the HOMO-LUMO gap of the system is high enough, and hence it is challenging to make the compound active in a chemical reaction. However, the concept of softness is reversed. A soft molecule has a relatively smaller HOMO-LUMO energy gap. The HOMO and LUMO energies indicate the ability to donate and accept an electron, respectively. So, we can compute the electron affinity and ionization potential by applying the LUMO and HOMO energies ($EA = -E_{LUMO}$ and $IP = -E_{HOMO}$). The HOMO and LUMO orbitals for the BNNT, BNNT/gas-molecule ($CH_2$, $SO_2$, $NH_3$, $H_2Se$, $CO_2$ and $CS_2$) complexes are plotted in Fig. 2. It is evident that for the BNNT/gas complex, the HOMO (nucleophile agent) is more localized over the atoms of a nanotube. The LUMO of the complex, which is an electrophile agent with an energy of about -0.62 eV, is localized upon the gas molecule, indicating very low electron affinity. Electrophilicity (ω) parameter gives an idea of the stabilization energy when the system gets saturated by electrons from the external reactive environment.

On the other hand, the reactive environment that donates an electron to the compound shows that it can contribute to a charge. So, it will behave as a nucleophilic compound in the reaction. As per the definition of different parameters, the more reactive nucleophile is characterized by a lower value of (ω). At the same time, higher values indicate an excellent electrophile system. Therefore, the high value of electrophilicity suggests a good electron receiver and the low electrophilicity means the compound will behave as a charge giver. Theoretically calculated, all of these chemical parameters are presented in Table 1. All these parameters together characterize the chemical behavior of the BNNT nanotube in pure and hybrid forms.

In addition to these chemical parameters, we have calculated the adsorption energy ($E_{add}$) and the charge transfer between the pure BNNT and the adsorbed gas molecules after forming the complex. In this study, structural optimizations and electronic properties of pristine BNNT and $CH_2$, $SO_2$, $NH_3$, $H_2Se$, $CO_2$ and $CS_2$ gas molecule adsorbed BNNTs are investigated as described by following the DFT method as described. For the adsorption of the gas molecule on the surface of pristine (5,5) BNNT, the Carbon or Sulphur atom of the gas molecule is bonded with a nitrogen atom in BNNT. To find the most favorable adsorption sites of BNNT, the gas molecules are placed at different positions on the BNNT surface with varying orientations concerning the surface with a restriction that all atoms of each gas molecule are pointed out into the boron (B) or nitrogen (N) atoms of the BNNTs. The optimized complex's bond length and adsorption energy analysis show two interactions between the BNNT and the gas molecules. These adsorptions can be classified into physisorption and chemisorption.

That physisorption is generally characterized by weak physical interactions such as Van der Waals forces and hydrogen bonding. In contrast, in the present study, chemisorption involves a chemical reaction via charge hybridization between surface molecules of BNNT and the adsorbed gas molecules, with relatively stronger bonds such as covalent bonding, electrostatic solid, and ionic bonding. The bond lengths for the BNNT nitrogen atom and the gas molecules of $CO_2$ (N-C), $SO_2$ (N-S), $CH_2$ (N-C), and $H_2Se$ (N-Se) are 3.25Å, 2.93Å, 1.55 Å and 3.27 Å respectively. On the other hand, the bond lengths between the BNNT boron molecule and different adsorbed gas molecules are $CS_2$ (B-C) and $NH_3$ (B-N), 3.68 Å and 1.71 Å, respectively. To avoid the effect of finite-length BNNT nanotube or edge effects, the unsaturated bonds of B or N are terminated by adding H atoms. Details of the bonds are presented in Table 1.

**Table 1: Different thermodynamic and chemical parameters of the BNNT (BN) and BNNT complex**

| **BNNT +Gas** | **Gap (eV)** | **$E_{add}$ (eV)** | **Bond Length** | **Mullikan Charge on the atoms of the adsorbed gas molecule** | | | | **Total Charge Transfer** | **From** | **To** |
|---|---|---|---|---|---|---|---|---|---|---|
| **BNNT** | 6.13 | | | | | | | | | |
| **$CO_2$** | 6.02 | 20.521 | 3.35 | C: 0.45 | O:-0.22 | O:-0.22 | | 0.011 | Gas | BN |
| **$SO_2$** | 2.17 | 21.45 | 2.93 | S: 0.71 | O: -0.38 | O: -0.39 | | -0.065 | BN | Gas |
| **$CS_2$** | 4.39 | 20.08 | 3.68 | C: -0.21 | S:0.10 | S:0.1 | | -0.01 | BN | Gas |
| **$CH_2$** | 5.77 | 23.53 | 1.55 | C:-0.36 | H: 0.13 | H: 0.14 | | -0.087 | BN | Gas |
| **$NH_3$** | 5.19 | 19.18 | 1.71 | N: -0.47 | H: 0.24 | H: 0.25 | H: 0.25 | 0.266 | Gas | BN |

| **$H_2Se$** | 6.01 | 19.32 | 3.27 | Sc: -0.08 | H: 0.09 | H: 0.09 | | 0.098 | Gas | BN |
|---|---|---|---|---|---|---|---|---|---|---|
| **BNNT +Gas** | **IP (eV)** | **EA (eV)** | **μ (eV)** | **η (eV)** | **ω (eV)** | **S $(eV)^{-1}$** | | | | |
| **BNNT** | 6.61 | 0.53 | -3.57 | 3.04 | 2.1 | 1.52 | | | | |
| **$CO_2$** | 6.68 | 4.51 | -5.6 | 1.09 | 14.43 | 0.55 | | | | |
| **$SO_2$** | 6.6 | 2.21 | -4.4 | 2.2 | 4.41 | 1.1 | | | | |
| **$CS_2$** | 6.28 | 0.38 | -3.33 | 2.96 | 1.88 | 1.48 | | | | |
| **$CH_2$** | 6.34 | 0.21 | -3.28 | 3.07 | 1.75 | 1.54 | | | | |
| **$NH_3$** | 6.56 | 0.35 | -3.46 | 3.11 | 1.92 | 1.56 | | | | |
| **$H_2Se$** | 6.59 | 0.52 | -3.55 | 3.04 | 2.08 | 1.52 | | | | |

To understand the electron delocalization between BNNT and the gas molecule that results in the interaction between the "BNNT/Gas molecule" system, natural bond population (NBO) analysis is performed, and the data is shown in Table 2. The formation of stable bonds between the nanotube surface and the adsorbed gas molecule is known to correspond to the electron density delocalization of bonding or lone pair and anti-bonding. In such a case, one unit behaves as an electron donor and the other as an acceptor. Details of the electron exchange can be understood from the data in Table 2. Electron density localization and EPS contour near the adsorbed gas molecules on the BNNT surface can be seen in Fig. 2.

**Table 2:** NBO analysis of BNNT-complex

| BNNT /gas | BNNT element | Gas Molecule | | |
|---|---|---|---|---|
| $CO_2$ | $N:2S^{0.68}2p^{2.41}$ | $C:2S^{0.34}2p^{1.17}3p^{0.01}$ | $O:2S^{0.87}2p^{2.36}$ | $O:2S^{0.87}2p^{2.36}$ |
| $CS_2$ | $B:2S^{1.36}p^{4.80}$ | $C:2S^{1.16}2p^{3.21}3p^{0.02}$ | $S:3S^{1.74}3p^{4.05}4S^{0.01}4p^{0.01}$ | $S:3S^{1.74}3p^{4.04}4S^{0.01}4p^{0.01}$ |
| $SO_2$ | $N:2S^{0.68}2p^{2.41}$ | $S:3S^{0.87}3p^{1.48}4S^{0.01}4p0^{.01}$ | $O:2S^{0.96}2p^{2.36}$ | $O:2S^{0.96}2p^{2.36}$ |
| $CH_2$ | $N:2S^{0.66}2^{p2.32}$ | $C:2S^{1.16}2p^{1.71}3p^{0.01}$ | $H: 1S^{0.39}$ | $H: 1S^{0.39}$ |
| | $B:2S^{0.24}p^{0.67}3p^{0.01}$ | | | |
| $NH_3$ | $B:2S^{0.21}2p^{0.71}3p^{0.01}$ | $N:2S^{0.53}p^{2.24}3p^{0.01}$ | $H: 1S^{0.29}$ | $H: 1S^{0.29}$ |
| | | | | $H: 1S^{0.3}$ |
| $H_2Se$ | $N:2S^{0.68}2^{p2.41}$ | $Se: 4S^{0.90}4p^{2.18}$ | $H: 1S^{0.46}$ | $H: 1S^{0.45}$ |

To understand the electronic charge, transfer between BNNT and the adsorbed gas molecule, the net charge transfer (Q) is defined as the charge difference between gas molecules adsorbed on the BNNT surfaces and isolated gas molecules after the formation of the BNNT-Gas complex. Adsorption energy and the charge transfer can be obtained by the equation as follows:

$$E_{Add} = E_{Complex} - E_{BNNT} - E_{Gas\,molecule};\ Q_{Net\ Transfer} = -Q_{Gas\ in\ complex} ----(5)$$

As per the above relation of charge transfer, since the total charge of the BNNT and gap molecules were in a neutral state, therefore the charge transfer from BNNT to gas molecule and vice versa can be estimated from the measurement of net Mulliken charge in the gas molecule after making the BNNT-Gas complex.

To introduce the same concept, in the present work, we define the stability of a complex in terms of the percentage change in adsorption energy after forming the complex as follows:

$$\%E_{add} = \left[\frac{E_{BNNT_Complex} - E_{BNNT} - E_{Gas_Molecule}}{E_{BNNT} + E_{Gas_Molecule}}\right] \times 100$$

Where $E_{BNNT_Complex}$, $E_{BNNT}$ and $E_{Gas_Molecule}$ are the energy of the BNNT complex, BNNT, and the gas molecule adsorbs. We defined these parameters based on previous reports on different systems [34-48]. As the results indicate, the energy gap of the nanotube has dramatically reduced after adsorption of the $CO_2$, $SO_2$, $CH_2$, $NH_3$, and $H_2Se$ molecules on the outer surface of the nanotube; so hardness values of BNNT-gas complexes have decreased in comparison to the pristine nanotube, which affirms more significant chemical activity of these complexes. The other relevant parameters of BNNT complexes are calculated in Table 1. We observe the charge distribution of the complexes, which shows the charge transfer before/after the interactions, depending on the electron affinity of the incoming complex. These are generated by the same DFT calculations in Gaussian, with the parameters discussed earlier. Fig. 2 shows that in orbital plots and different contour presentations along the longitudinal direction and cross-sectional view, the gas molecules are well under the interaction of the bare BNNT. Therefore, BNNT is a potential candidate as pollutants gas molecule sensor.

In the end, we have studied the IR spectrum of the BNNT and BNNT+$CO_2$ complex (shown in Fig. 3) to confirm the absorption of $CO_2$ molecule over the surface of BNNT. From the IIR spectrum of BNNT and its complex, we found that the dominating frequency of the bare BNNT (1520 $cm^{-1}$) shifted towards the higher side (1525 $cm^{-1}$) after the adsorption of $CO_2$. From the IR spectrum, it is found that the dominating frequency mode of bare $CO_2$ (2360 $cm^{-1}$) shifted to 2353 $cm^{-1}$ due to interaction with BNNT. We have yet to calculate the IR spectrum of other BNNT + Gas molecule complexes. But the example of $CO_2$ makes us believe that a similar effect on the IR spectrum can be found after the adsorption of the gas molecules on the BNNT surface.

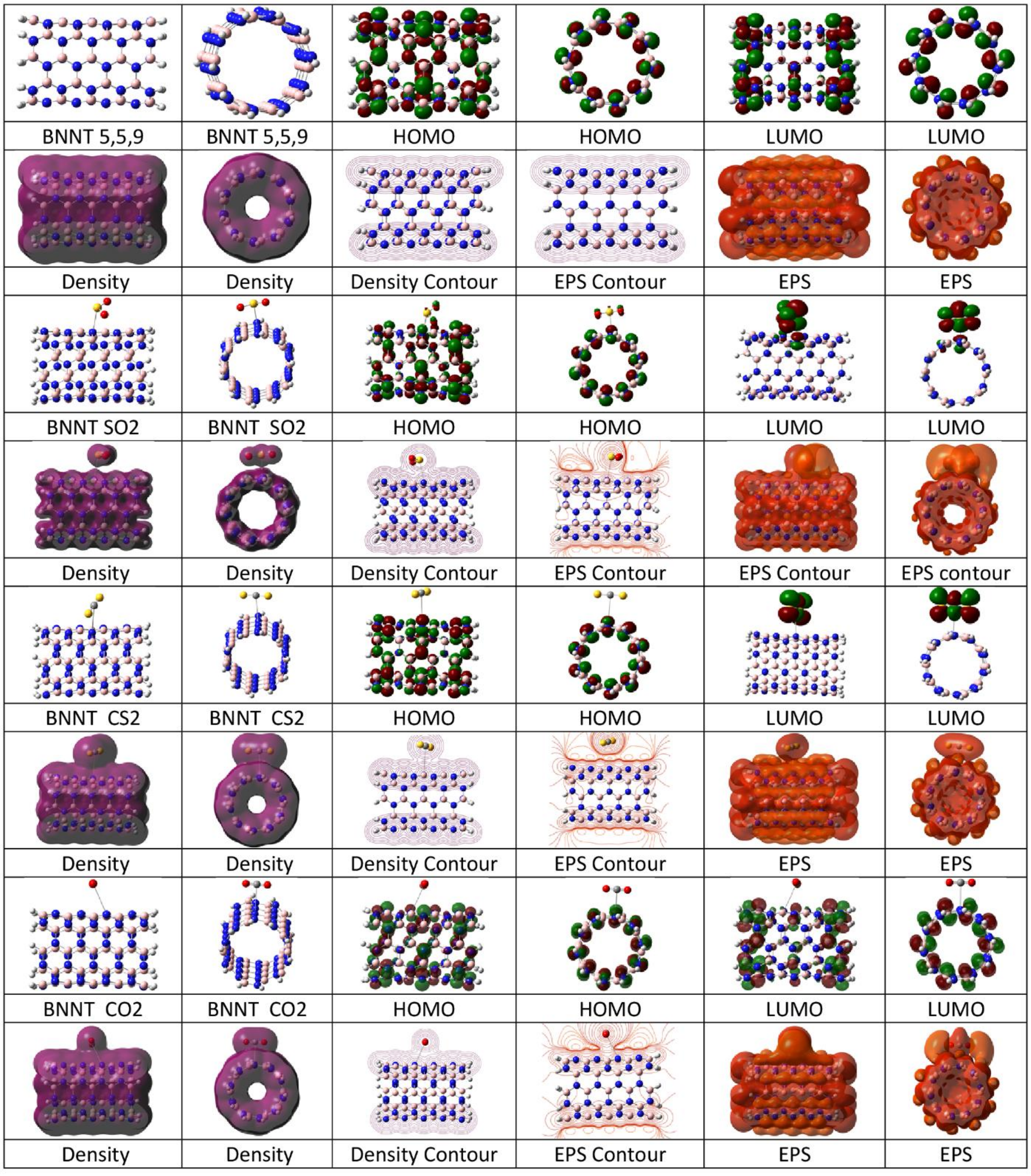

Figure 2: Optimized structure, HOMO-LUMO, electron density, and the electrostatic potential surface of pure and BNNT-gas complexes for $SO_2$, $CS_2$, and $CO_2$.

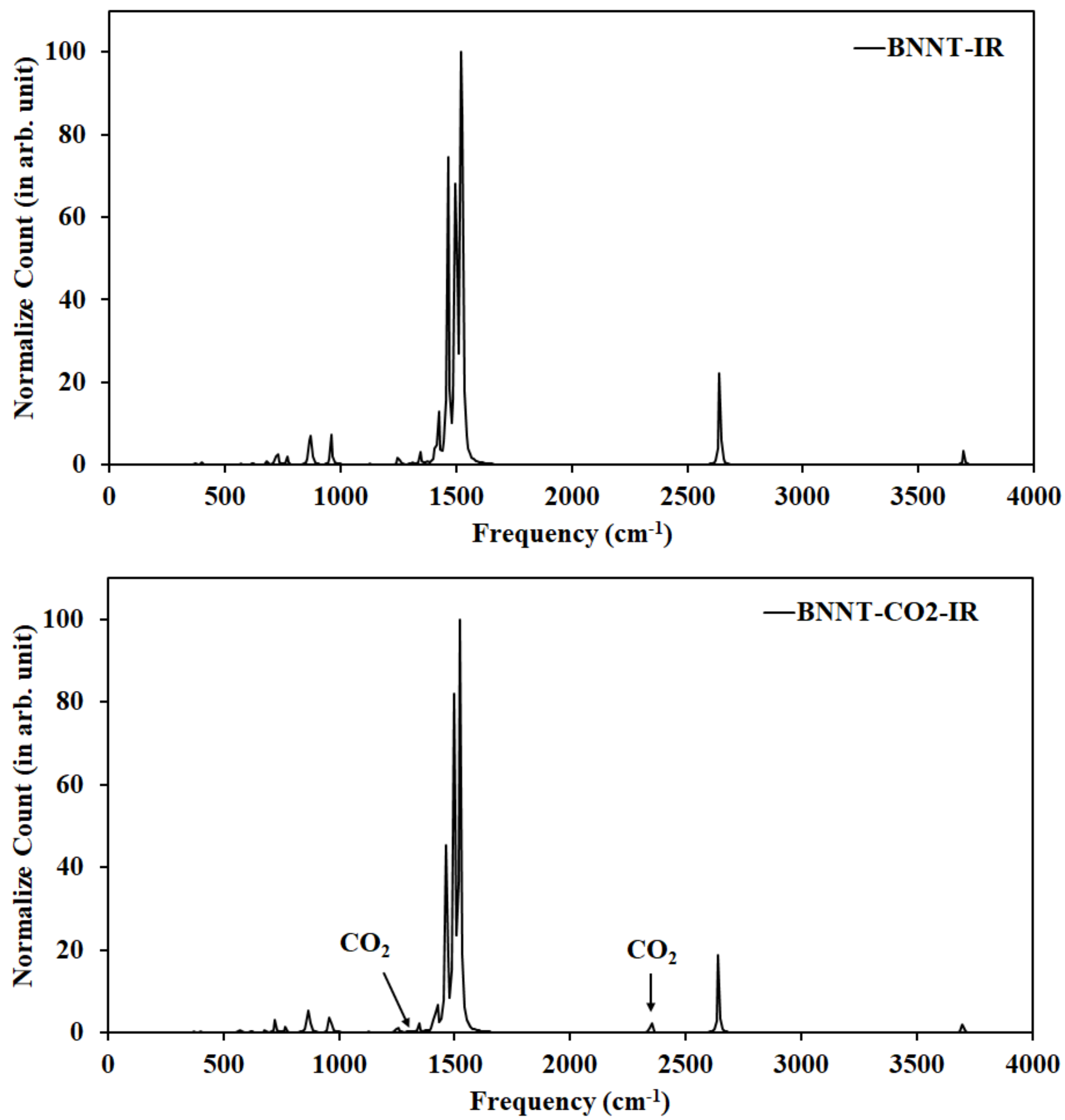

Figure 3. The IR spectrum of pure BNNT and BNNT+$CO_2$ complex

**CONCLUSION**

In the report, we set out to determine the suitability of BN nanotubes for applications in pollutant sensors and absorption. We performed DFT calculations and studied the chemical properties of the nanotube to understand the efficiency of the BNNT as a gas absorbent.

In this work, we carried out a linear combination of atomic orbital density functional theory (DFT) together with a spin-polarized generalized gradient approximation (GGA) to study the interaction of different gas molecules, such as $CH_2$, $SO_2$, $NH_3$, $H_2Se$, $CO_2$ and $CS_2$ molecules with the outer surface of pristine BNNTs. The armchair (5,5) chirality is investigated. These pollutant gas molecules interact weakly with pristine BNNTs through van der Waals-like interactions. It was concluded that the most suitable system as far as the adsorption energy is concerned depends on the gas molecules. The gasp molecules $CH_2$, $SO_2$, $NH_3$, $H_2Se$, $CO_2$ and $CS_2$ are adsorbed on the BNNT nanotube with N, B, N, B+N, B, and N, respectively. We have noted that the BNNT–gas complex presented chemisorption states far superior to those observed for carbon nanotubes. Compared to carbon nanotubes, the better quality of BNNTs, as potential candidates as gas adsorbents, was confirmed through Mulliken charge analysis. The IR spectrum of the BNNT and BNNT+$CO_2$ complex confirms that the BNNT adsorbs the $CO_2$ gas molecule. So, this could be effective for other gas molecules also. These facts and advantages, such as excellent chemical stability and BNNTs, make us consider them promising structures for detecting, capturing, and adsorbing different pollutant gas molecules. Therefore, we have shown that these systems are tools of significant importance for developing an environmentally desired mechanism of pollutant gas arresting. Overall, while more research is needed to fully understand the potential of BNNTs as gas molecule absorbents, the unique properties of these materials make them promising candidates for a wide range of applications, including gas separation, purification, and storage

**CRediT authorship contribution statement**

The manuscript, data simulation, literature survey, and corrections were collaborative efforts involving all authors. Debashis Bandyopadhyay took the lead in conceptualization, planning, and finalizing the corrected draft.

**Declaration of Competing Interest**

The authors declare that they have no conflicts of interest to this work.